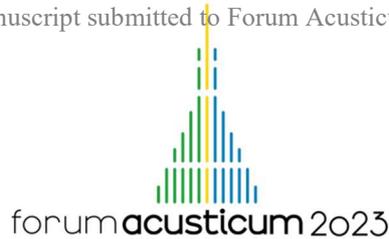

# COMPUTATIONALLY-EFFICIENT AND PERCEPTUALLY-MOTIVATED RENDERING OF DIFFUSE REFLECTIONS IN ROOM ACOUSTICS SIMULATION

**Stephan D. Ewert[1]\***    **Nico Gößling[1]**    **Oliver Buttler[1]**
**Steven van de Par[1]**    **Hongmei Hu[1]**

[1] Medizinische Physik und Akustik and Cluster of Excellence Hearing4all, Universität Oldenburg, 26111 Oldenburg, Germany

## ABSTRACT

Geometrical acoustics is well suited for simulating room reverberation in interactive real-time applications. While the image source model (ISM) is exceptionally fast, the restriction to specular reflections impacts its perceptual plausibility. To account for diffuse late reverberation, hybrid approaches have been proposed, e.g., using a feedback delay network (FDN) in combination with the ISM. Here, a computationally-efficient, digital-filter approach is suggested to account for effects of non-specular reflections in the ISM and to couple scattered sound into a diffuse reverberation model using a spatially rendered FDN. Depending on the scattering coefficient of a room boundary, energy of each image source is split into a specular and a scattered part which is added to the diffuse sound field. Temporal effects as observed for an infinite ideal diffuse (Lambertian) reflector are simulated using cascaded all-pass filters. Effects of scattering and multiple (inter-) reflections caused by larger geometric disturbances at walls and by objects in the room are accounted for in a highly simplified manner. Using a single parameter to quantify deviations from an empty shoebox room, each reflection is temporally smeared using cascaded all-pass filters. The proposed method was perceptually evaluated against dummy head recordings of real rooms.

---

\***Corresponding author**: *stephan.ewert@uni-oldenburg.de.*

## 1. INTRODUCTION

Room acoustics simulation has various applications ranging from room-acoustical planning, hearing research, rehabilitation and training, to virtual reality applications and artificial reverberation in video games (e.g., [1, 2]). Accordingly, the research field has gained rising popularity [3-5] with increasing demand for real-time applicability and interactive virtual acoustic environments, e.g., [6-10]. These developments require computationally efficient simulation techniques. Geometrical acoustics (GA) is frequently used in this context and well suited to model direct sound and early reflections, assuming ray-like sound propagation and specular reflections.

However, sound scattering plays an important role in room acoustics and related phenomena such as edge diffraction can be integrated into GA by considering rays with "bended" propagation paths (e.g., [11-13]). Such paths involve path nodes at edges in the environment, e.g., at room or building corners, openings, or smaller objects such as tables. Computationally highly-efficient, digital filter-based diffraction solutions for GA were proposed in [14-16], offering a physically-based simulation of diffraction from arbitrary infinite and finite wedges (and objects composed thereof). For complex geometries with many edges and objects, however, diffraction path tracing and detailed diffraction modelling become costly. Thus, more simplified approximations might be helpful. One approach for objects using simple parametric filters estimated by machine learning has been studied in [17].

Besides the above described effects of scattering at large structures and objects, scattering also occurs at boundaries where the effect of small-scale geometric or spatial impedance variations can be considered as surface "roughness" and may be effectively described by the frequency-dependent scattering coefficient [18]. In contrast to GA specular reflections at boundaries as accounted for by the image source model (ISM; [19, 20]), sound reflections in real rooms typically involve diffuse reflections resulting from scattering at rough surfaces. Such diffuse reflections are dominating after the third reflection order [21]. Moreover, the effect of diffuse reflections was shown to be audible [22] and should thus be considered in virtual acoustic environments. For example, an extension of the ISM by incoherent reflections from rough room surfaces was suggested in [23]. In, e.g., [24, 25] the ISM was used for low-order specular reflections, while diffuse (later) reflections were modelled in a simplified manner using a feedback delay network (FDN; [26]) with room-dependent parameters and spatial rendering. However, with such a processing scheme, the transition from purely specular to purely diffuse reflections occurs at a distinct, pre-defined reflection order.

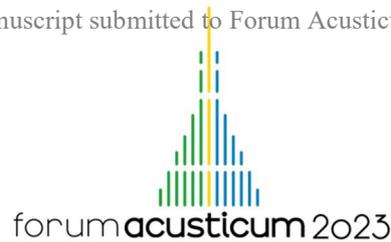

Taken together, computationally-efficient approximations for scattering at surfaces and for sound paths involving partly obstructing larger scale geometric structures and objects in a room are desirable, especially for real-time virtual acoustics. Parametric digital filters appear particularly suited to achieve the desired low computational complexity. So far, such filter solutions exist for scattering phenomena such as discrete diffracted sound paths at edges and objects (e.g., [16]).

Here, a strongly simplified and computationally highly efficient approach to approximate the global effects of sound scattering and resulting diffuse reflections for room acoustic simulation is suggested, aiming at perceptual plausibility.

Depending on the frequency-dependent scattering coefficient [18] of a surface, we assume ideal diffuse (Lambertian) reflections (e.g., [27, 28]) in addition to the GA specular reflection in the ISM. We propose splitting the energy of each image source (IS) into a specular and non-specular (diffuse) part using IS decomposition filters. The non-specular reflected part of each IS is then fed into an FDN serving as diffuse reverberation model ([24, 25]). As a consequence, the suggested approach realistically generates (partly) diffuse reflections for all ISs orders. In the time domain, ideal diffuse reflections from a wall surface lead to a temporally distributed summed reflection, given that each surface point contributes with different travel time and attenuation. To approximate this effect, we propose the use of highly-efficient cascaded all-pass filters to temporally "smear" the non-specular parts of the ISs. The accompanying spatial spread is accounted for by mapping the diffuse reflections to the according directions of virtual reverberation sources used in the FDN spatialization [29, 30].

Complementing the existing physically-based filter solution for discrete diffracted sound paths at edges and objects of [16], we suggest to account for the net effect of scattering and multiple (inter-) reflections at larger geometric structures and by objects in the room in a highly simplified manner: We use a single parameter to quantify geometric deviations from an ideal empty shoebox room. Using this parameter, each IS reflection is temporally distributed using cascaded all-pass filters. The proposed method was perceptually evaluated against dummy head recordings of real rooms using a subset of the spatial audio quality inventory [31].

The suggested algorithms are freely available in the framework of the room acoustics simulator (RAZR; *www.razrengine.com*).

## 2. SURFACE SCATTERING

In order to characterize the effects of surface scattering, for simplicity, a single rigid infinite wall is assumed with the source and receiver located in front of the wall at the same point at a distance $R$. Following Fermat's principle, the shortest reflection path corresponds to the specular reflection path, which is in this case perpendicular to the wall with total distance of $2R$.

Further assuming ideal (Lambertian) diffuse reflections (e.g., [27, 28]), "non-specular" reflections from each point on the wall surface additionally arrive at the receiver. These diffuse reflections are frequency-dependent and can be expressed using the scattering coefficient [18], depending on the size of the wall irregularities and roughness in relation to the wavelength of sound. Typically, scattering increases towards high frequencies, resulting in different spectra for the diffusely and specularly reflected energy. The contribution from each point on the wall to the diffusively reflected sound arrives with a delay relative to the specular reflection at the receiver, caused by longer path lengths.

To quantify the resulting temporal spread of the diffuse reflection, a monotonic temporal decay can be derived with the above assumptions (not shown here) and expressed in dB, similar to an energy decay curve (EDC). The effect of surface scattering can conceptually be interpreted as a local energy decay, following each specular reflection. A local decay time $T_s$ can be derived, for simplicity assuming an exponential energy decay. Roughness, bumpiness, or impedance variations along the wall's surface would lead to a noisy phase of the diffuse reflection.

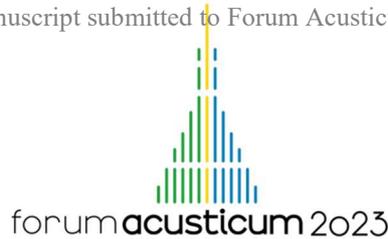

We propose to efficiently approximate the main properties of the decay using all-pass filter structures [32-34], originally suggested for artificial room reverberation. To achieve a pulse density that is sufficiently high for the human listener, Schroeder [32] connected these all-pass reverberators in series to form an all-pass cascade (APC). The here suggested decay filter (with "colorless" long-term spectrum) uses a series of four all-pass filters of the form

$$H_i(z) = \frac{g_i + z^{-\tau_i}}{1 + g_i z^{-\tau_i}}, \qquad i = 1 \dots 4, \qquad (1)$$

with

$$\tau_i = \frac{\tau_s f_s}{\eta^{i-1}}, \qquad \tau_s = \frac{T_s \log|g^{-1}|}{3}, \qquad (2)$$

where all gains $g_i = \sqrt{2}^{-1}$ are fixed and identical, $f_s$ is the sampling frequency, and the delay ratio of the different all-pass filters depends on the parameter $\eta = \pi$, to achieve a (nearly) non-commensurate ratio [34]. The delays $\tau_i$ are rounded to the closest integer sample value. The local decay time $T_s$ can be derived based on the distances to the reflecting wall as outlined above. Alternatively, for a given room, the mean distance to all walls, or the mean free path length may be used, neglecting the (time-variant) source and receiver positions, and resulting in a time-invariant APC. The number of four all-pass filters was chosen to keep computational costs low, while achieving a sufficiently high pulse density for typical room dimensions (see [32]).

In addition to the above described temporal spread, surface scattering causes a spatial spread of the diffuse reflection with contributions impinging from the solid angle covered by the reflecting surface. The spatial spread is represented in a simplified manner by mapping the output of the all-pass filter to those spatially distributed virtual reverberation sources of the FDN spatialization [29, 30] which represent the reflecting surface.

## 2.1 Decomposition filters

Specularly reflected and scattered sound are implemented for each image source in an ISM using parametric decomposition filters based on the frequency-dependent scattering coefficient [18]. The temporal spread of the diffuse reflections is introduced by applying the above described APC, before feeding the diffuse part into a diffuse reverberation model. Here, this scheme was implemented in ([24, 25]) using a spatially-mapped FDN as diffuse reverberation model. The mapping of the FDN output to spatially evenly-distributed virtual reverberation sources (VRS; [25, 29]) around the listener is utilized and diffuse

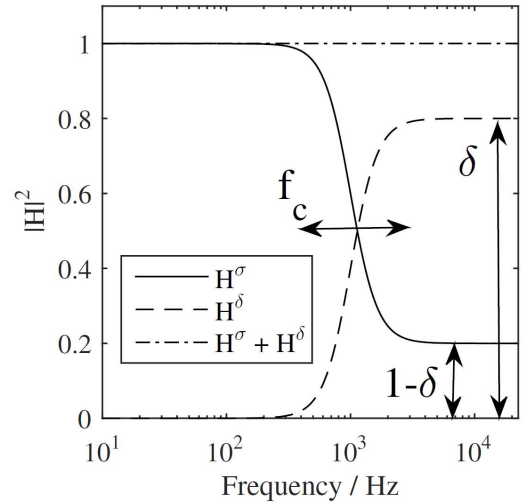

**Figure 1.** Parameterized image source decomposition using two power-complementary filters $H^\sigma$ and $H^\delta$ for the specular and diffuse reflection, respectively.

reflections are assigned to VRS representing the direction of the reflecting surface.

Two power-complementary IS decomposition filters (second-order, low-shelving filter for the specular part, see Fig. 1) were implemented assuming that the scattering coefficient $\delta$ is generally low for low frequencies and high for high frequencies (see, e.g., [28]). The transition between the low-frequency specular part $H^\sigma$ and the high-frequency diffuse part $H^\delta$ is specified by the cut-off frequency $f_c$. With this simplified filter design only two parameters, $\delta$ and $f_c$ are used to the approximate measured scattering coefficients for each wall.

## 3. OBJECT SCATTERING

For the effective simulation of sound scattering due to interior objects such as furniture or larger scale irregularities at room boundaries, a further APC-based approach is suggested. The simplified effect of such geometric deviations from an empty shoebox is that no single specular or diffuse reflection from a wall with one associated single delay (and surface scattering tail) is observed, but multiple scattered (inter-) reflections between surfaces of the objects occur on the path, which arrive at different delays at the receiver. For the temporal distribution of the summed reflections from multiple objects at the receiver, an envelope shape similar to a gamma distribution function can be assumed [23].

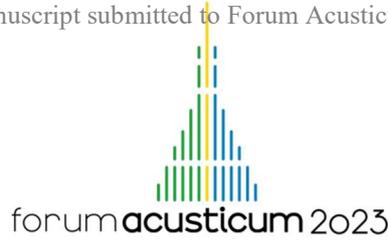

A single-valued geometric deviation parameter is estimated, referred to as $\zeta$, expressing the deviation from an (ideal) empty shoebox room. For an empty shoebox, the geometric deviation is defined as $\zeta = 0$. A typical office room with some shelves at the walls and desks would be characterized by an averaged geometric deviation of about 0.05 to 0.1, depending on the size of the shelves relative to the room dimensions. A storage room filled with shelves and goods would have a geometric deviation of 0.4 to 0.6.

To efficiently approximate the desired temporal distribution of summed reflections, again a series of four all-pass filters as in Eqn. (1) is used. Here, the gain factors $g_i$ and delays $\tau_i$ are slightly altered to achieve the desired gamma-distribution-like envelope:

$$\boldsymbol{g} = \left( \sqrt{2}^{-1}, \sqrt{2}^{-1}, \sqrt{2}^{-2}, \sqrt{2}^{-3} \right), \tag{3}$$

$$\boldsymbol{\tau} = \left( \frac{\tau_0 f_s}{\eta^2}, \frac{\tau_0 f_s}{\eta^2}, \frac{\tau_0 f_s}{\eta}, \tau_0 f_s \right), \quad \tau_0 = \frac{\gamma}{\sum_{k=0}^{3} \eta^{-k}}. \tag{4}$$

The delays $\tau_i$ are rounded to integer sample values. $\gamma$ is the desired group delay of the APC and is calculated from the (specular) reflected overall path length $d$ (based on the shoebox room boundaries), the speed of sound $c$ and the geometric deviation parameter $\zeta$:

$$\gamma = \zeta \frac{d}{c}. \tag{5}$$

For $\zeta > 0$, the specular reflections in the ISM are filtered with a specific APC. Without such a filter, particularly for large rooms, where individual early specular reflections are sparse, pure specular reflections (including a surface scattering tail) in the ISM might result in an unrealistic "crackling" impression for transient sounds. In this case, the simplified effect of object scattering is assumed to be audible, even for a small geometric deviation parameter of $\zeta = 0.05$ (5%).

## 4. PERCEPTUAL EVALUATION

The suggested object scattering was evaluated by comparing measured and simulated (implemented in RAZR, see [24, 25] for further details) binaural room impulse responses (BRIRs) for several rooms. A 3rd-order ISM was used and 12 channels in the FDN. Perceptual ratings of a subset of attributes from the spatial audio quality inventory (SAQI; [31]) were used. In order to better account for the audible effects specifically associated with scattering, two additional attributes *Distortion* and *Flutter echoes* were introduced, based on informal listening tests.

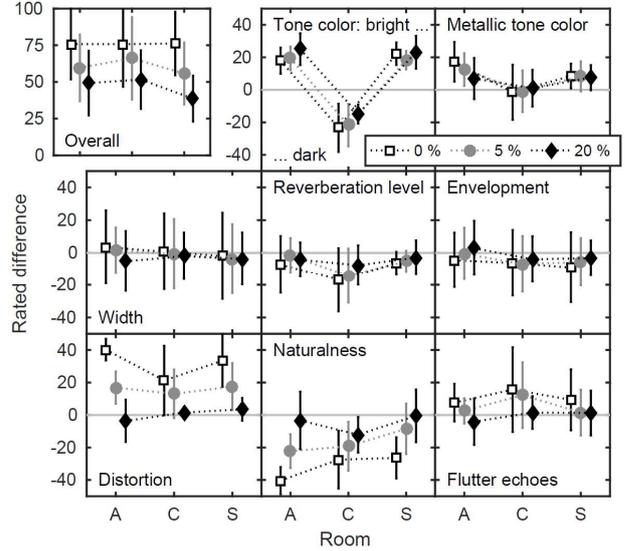

**Figure 2**. Average rated differences between measured and synthesized BRIRs (convolved with a pink pulse) for different sound attributes (see panels), plotted against rooms. Error bars indicate inter-subject standard deviations. Depending on the attribute, ordinate scales ranged from, e.g., "less pronounced" to "more pronounced" or semantically fitting descriptors.

### 4.1 Participants

Eight normal-hearing participants (all male) aged 29 to 46 years participated in the experiment. All listeners were working in the field of (virtual) acoustics or hearing research and were experienced in performing psychoacoustic experiments. They were therefore considered expert listeners.

### 4.2 Rooms and test signals

Three rooms, A (large, Aula Carolina, Aachen), C (medium, corridor, Oldenburg), and S (medium, seminar room, Oldenburg) were used (for a detailed description see [24, 25]). For each room, BRIRs for relative geometric deviations of 0% (i.e., no scattering), 5%, and 20% were created. Head-related impulse responses of the MK2 artificial head by Cortex were used (like for the recordings in room C and S). As source signal, a delta pulse was used that was filtered to obtain a pink-spectrum-shaped frequency response. All stimuli were 2 s in duration. The participants were seated in a sound-attenuating listening booth and listened using Sennheiser HD 650 headphones

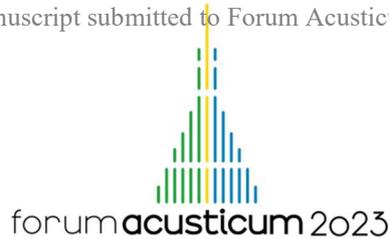

(equalized transfer function), driven by an RME ADI-2 soundcard at a sampling rate of 44.1 kHz.

### 4.3 Results and discussion

The results for the rated perceived differences between the measured (reference) and simulated conditions, averaged over listeners are shown in Fig. 2. Ratings are compared for the different geometric deviations (0%-20%). The overall differences (upper left panel) to the measured reference decreases with an increasing geometric deviation parameter. This indicates that the suggested scattering approximation leads to a higher perceived similarity between the recording and the simulation. For most attributes (other panels), perceived differences are generally small. A distinct effect of the scattering simulation is observed for the attributes *Distortion*, *Naturalness*, and *Flutter echoes*. Differences to the measured conditions are the smaller, the larger the geometric deviation. For *Tone color*, differences in both directions (bright, dark) are perceived indicating overall spectral differences between the recording and the simulation.

### 5. SUMMARY AND CONCLUSIONS

Two strongly simplified digital-filter approximations were suggested to account for the effects associated with scattered reflections at surfaces and objects for GA-based room acoustics simulation. Computationally highly-efficient all-pass cascades are used to mimic effects of scattering for each specular reflection in an image source model. The parameters of the APCs are chosen to either account for a monotonically decaying temporal spread caused by surface scattering or for a gamma-distribution-like envelope of the spread caused by multiple scattered (inter-) reflections from objects in the room.

The suggested surface scattering approach transfers scattered sound energy for each reflection in the ISM to a diffuse reverberation model using decomposition filters based on the frequency-dependent scattering coefficient. The spatial spread is accounted for by mapping the scattered reflections to spatially distributed virtual reverberation sources used in the diffuse reverberation model, according to the solid angle covered by the considered surface.

The suggested object scattering, using a single parameter to specify the geometric deviation from an empty shoebox room, avoids unrealistic sparse specular reflections, particularly for large rooms, without the necessity to model a large number of individual reflecting surfaces. Room acoustics simulations with object scattering were perceptually rated more similar to recordings in real rooms. Both suggested approaches are highly suited for real-time applications.

Future research should extend the perceptual evaluation to other source signals. Machine learning might be applicable to estimate the geometric deviation parameter from room models or camera images.

### 6. ACKNOWLEDGMENTS

This work was supported by the Deutsche Forschungsgemeinschaft, DFG – Project-ID 352015383 – SFB 1330 C5 and DFG SPP Audictive – Project-ID 444827755

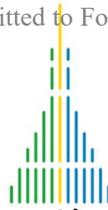